# Sketch-based Querying of Distributed Sliding-Window Data Streams

Odysseas Papapetrou, Minos Garofalakis, Antonios Deligiannakis
Technical University of Crete
{papapetrou,minos,adeli}@softnet.tuc.gr

## ABSTRACT

While traditional data-management systems focus on evaluating single, ad-hoc queries over static data sets in a centralized setting, several emerging applications require (possibly, continuous) answers to queries on *dynamic* data that is widely *distributed* and constantly updated. Furthermore, such query answers often need to discount data that is "stale", and operate solely on a *sliding window* of recent data arrivals (e.g., data updates occurring over the last 24 hours). Such *distributed data streaming* applications mandate novel algorithmic solutions that are both time- and space-efficient (to manage high-speed data streams), and also communication-efficient (to deal with physical data distribution). In this paper, we consider the problem of complex query answering over distributed, high-dimensional data streams in the sliding-window model. We introduce a novel sketching technique (termed *ECM-sketch*) that allows effective summarization of streaming data over both time-based and count-based sliding windows with probabilistic accuracy guarantees. Our sketch structure enables point as well as inner-product queries, and can be employed to address a broad range of problems, such as maintaining frequency statistics, finding heavy hitters, and computing quantiles in the sliding-window model. Focusing on distributed environments, we demonstrate how ECM-sketches of individual, local streams can be composed to generate a (low-error) ECM-sketch summary of the order-preserving aggregation of all streams; furthermore, we show how ECM-sketches can be exploited for continuous monitoring of sliding-window queries over distributed streams. Our extensive experimental study with two real-life data sets validates our theoretical claims and verifies the effectiveness of our techniques. To the best of our knowledge, ours is the first work to address efficient, guaranteed-error complex query answering over distributed data streams in the sliding-window model.

## 1. INTRODUCTION

The ability to process, in real time, continuous high-volume *streams* of data is a common requirement in many emerging application environments. Examples of such applications include, sensor networks, financial data trackers, and intrusion-detection systems. As a result, in recent years, we have seen a flurry of activity in the area of *data-stream processing*. Unlike conventional database query processing that requires several passes over a static, archived data image, data-stream processing algorithms often rely on building concise, approximate (yet, accurate) *sketch synopses* of the input streams in real time (i.e., in one pass over the streaming data). Such sketch structures typically require *small space and update time* (both significantly sublinear in the size of the data), and can be used to provide *approximate query answers* with guarantees on the quality of the approximation. These answers can be more than sufficient for typical exploratory analysis of massive data, where the goal is to detect interesting statistical behavior and patterns rather than obtain answers that are precise to the last decimal. Large-scale stream processing applications are also inherently *distributed*, with several remote sites observing their local stream(s) and exchanging information through a communication network. This distribution of the data naturally imposes critical *communication-efficiency* requirements that prohibit naïve solutions that centralize all the data, due to its massive volume and/or the high cost of communication (e.g., in sensornets). Communication efficiency is particularly important for *distributed event-monitoring* scenarios (e.g., monitoring sensor or IP networks), where the goal is real-time tracking of distributed measurements and events, rather than one-shot answers to sporadic queries [25].

Several query models for streaming data have been explored over the past decade. Streaming data items naturally carry a notion of "time", and, in many applications, it is important to be able to downgrade the importance (or, weight) of older items; for instance, in the statistical analysis of trends or patterns in financial data streams, data that is more than a few months old might be considered "stale" and irrelevant. Various *time-decay models* for querying streaming data have been proposed in the literature, mostly differentiating on the relation of an item's weight to its age (e.g., exponential or polynomial decay [6]). The *sliding-window* model [12] is one of the most prominent and intuitive time-decay models that considers only a window of the most recent items seen in the stream thus far (i.e., items outside the window are "aged out" or given a weight of zero). The window itself can be either *time-based* (i.e., items seen in the last $N$ time units) or *count-based* (i.e., the last $N$ items). Several algorithms have been proposed for maintaining different types of statistics over sliding-window data streams while requiring time and space that is significantly sublinear (typically, poly-logarithmic) in the window size $N$ [12, 15, 24, 26]. Still, the bulk of existing work on the sliding-window model has focused on tracking basic counts and other simple aggregates (e.g., sums) over one-dimensional streams in a centralized setting. Some recent work has also considered the case of distributed data, however, no existing techniques can handle flexible, complex aggregate queries over rapid, high-dimensional distributed data streams, e.g., with each dimension corresponding to the frequency of a distinct key in the stream.

*Example:* Recent work on effective network-monitoring systems (e.g., for detecting DDoS attacks or network-wide anomalies in large-scale IP networks) has stressed the importance of an efficient *distributed-triggering* functionality [20, 22, 18, 17]. In their early work, Jain et al. [20] discuss a generic distributed attack-detection scheme relying on the ability to maintain frequency statistics for high-dimensional data over sliding windows. In particular, each node (e.g., a network router implementing Cisco's Netflow protocol, a wireless access point, or a peer in a P2P network) maintains a sliding-window count of all observed messages for each target IP address. If this count exceeds a pre-determined threshold, which is determined based on the capacity of the target machine (possibly expressing the fair share of each client to the target machine), an event is triggered to a central coordinator as a warning of possible overloading. The coordinator then collects network-wide statistics to monitor overloaded nodes or abnormal behavior. More recent efforts have focused on different variants and extensions of this basic scheme, often requiring more extensive data/statistics collection and more sophisticated analyses [18, 17]. (Note that such data collection mechanisms are supported by commercial products, such as the Cisco Netflow Collection Engine solution.)

The ability to efficiently summarize high-dimensional data over sliding windows is obviously crucial to such network-monitoring schemes, given the tremendous volume of network-data streams and their massive domain sizes (e.g., $2^{48}$ for IPv6 addresses). This raises a critical need for synopsis data structures that can compactly capture accurate frequency statistics for a vast domain space over sliding windows. Furthermore, to enable the coordinator to aggregate data coming from different nodes (a requirement for detecting DDoS attacks), we need to be able to *compose* individually constructed synopses to a single synopsis which can capture the global state of the network and help isolate network-wide abnormalities. Thus, we are faced with the difficult challenge of designing effective, composable synopses that can support potentially complex sliding-window analysis queries over massive, distributed network-data streams. □

Note that similar requirements are frequently observed in other domains, e.g., for identifying misbehaving nodes in large wireless networks, for training of classifiers with distributed training data that expires over time, and for ranking products in a cloud-based e-shop, based on the number of recent visits of each product.

**Our Contributions.** In this paper, we consider the problem of answering potentially complex queries over distributed, high-dimensional data streams in the sliding-window model. Our contributions can be summarized as follows.

• **ECM-Sketches for Sliding-Window Streams.** We introduce a novel sketch synopsis (termed *ECM-sketch*) that allows effective summarization of streaming data over both time-based and count-based sliding windows with probabilistic accuracy guarantees. In a nutshell, our ECM-sketch combines the well-known Count-Min sketch structure [10] for conventional streams with state-of-the-art tools for sliding-window statistics. The end result is a sliding-window sketch synopsis that can provide provable, guaranteed-error performance for point, as well as inner-product, queries, and can be employed to address a broad range of problems, such as maintaining frequency statistics, finding heavy hitters, and computing quantiles in the sliding-window model.

• **Time-based Sliding Windows over Distributed Streams.** Focusing on distributed environments, we demonstrate how ECM-sketches summarizing time-based sliding windows of individual, local streams can be composed to generate a guaranteed-error ECM-sketch synopsis of the order-preserving aggregation of all streams. While conventional Count-Min sketches are trivially composable, composing ECM-sketches is more challenging, since it requires the composition of the sliding-window statistics maintained in the sketch. Compared to earlier work on composable, randomized sliding-window statistics [27, 15], our sliding window approximation technique is completely deterministic and is much more space efficient (with a linear rather than a quadratic dependence on the approximation error). This increased efficiency comes at the cost of a slight inflation of the worst-case error guarantee due to composition. Furthermore, we demonstrate how our ECM-sketches can be exploited in the context of the geometric framework of Sharfman et al. [25] for continuous monitoring of sliding-window queries over distributed streams.

• **Experimental Study and Validation.** We perform a thorough experimental evaluation of our techniques using two real-life data sets, in both centralized and distributed settings. The results of our study verify the efficiency and effectiveness of our ECM-sketch synopses in a variety of applications, and expose interesting functional trade-offs. When compared to algorithms based on randomized sliding window synopses – which are the only ones that were considered for composition up to now – ECM-sketches reduce the memory and computational requirements by at least one order of magnitude with a very small loss in accuracy. Similar savings apply to the network requirements.

## 2. RELATED WORK

**Centralized and Distributed Data Streams.** Most prior work on data-stream processing has focused on developing space-efficient, one-pass algorithms for performing a wide range of *centralized, one-shot computations* on massive data streams; examples include computing quantiles [16], estimating distinct values [14], counting frequent elements (i.e., "heavy hitters") [5, 9], and estimating join sizes and stream norms [1, 10]. Out of these efforts, flexible, general-purpose sketch summaries, such as the AMS [1] and the Count-Min [10] sketch have found wide applicability in a broad range of stream-processing scenarios. More recent efforts have also concentrated on *distributed-stream* processing, proposing communication-efficient streaming tools for handling a number of query tasks, including distributed tracking of simple aggregates [23], quantiles [8], and join aggregates [7], as well as monitoring distributed threshold conditions [25]. All the above-referenced works assume a traditional, "full-history" data stream and do not address the issues specific to the sliding-window model.

**Sliding-Window Stream Queries.** As mentioned earlier, the bulk of existing work on the sliding-window model has focused on algorithms for maintaining simple statistics, such as basic counts and sums, in space and time that is significantly sub-linear (typically, poly-logarithmic) in the sliding-window size $N$. *Exponential histograms* [12] are a state-of-the-art deterministic technique for maintaining $\epsilon$-approximate counts and sums over sliding windows, using $O(\frac{1}{\epsilon} \log^2 N)$ space. *Deterministic waves* [15] solve the same basic counting/summation problem with the same space complexity as exponential histograms, but improve the worst-case update time complexity to $O(1)$; on the other hand, *randomized waves* [15] rely on randomization through hashing to track *duplicate-insensitive* counts (i.e., COUNT-DISTINCT aggregates) over sliding windows. While randomized waves can be easily composed (in distributed settings), they also come with an increased space requirement of $O(\frac{\log(1/\delta)}{\epsilon^2} \log^2 N)$, where $\delta$ is a small probability of failure. Xu et al. [27] describe a randomized, sampling-based synopsis, very similar to randomized waves, for tracking sliding-window counts and sums with out-of-order arrivals (e.g., due to network delays) in a distributed setting. As with randomized waves, their space requirements are also quadratic in the inverse approximation error;

furthermore, their approach requires knowledge of the maximum number of elements in any sliding window (to set up the synopsis data structure), which could be problematic in dynamic, widely-distributed environments. Cormode et al. [11] also propose randomized techniques for handling out-of-order arrivals for tracking duplicate-insensitive sliding-window aggregates. To address the high cost associated with randomized data structures, Busch and Tirthapura propose a deterministic structure for handling out-of-order arrivals in sliding windows [3]. Similar to the other deterministic structures, this structure also does not allow composition and focuses only on basic counts and sums. Finally, Chan et al. [4] investigate continuous monitoring of exponential-histogram aggregates over distributed sliding windows. The main contribution of their work lies in the efficient scheduling of the propagation of the local exponential-histogram summaries to a coordinator, without violating prescribed accuracy guarantees.

Going beyond counts, sums, and simple aggregates, there is surprisingly little work in the more general problem of maintaining general, frequency-distribution synopses over high-dimensional streaming data in the sliding-window model. Hung and Ting [19] and Dimitropoulos et al. [13] propose synopses based on Count-Min sketches for tracking heavy hitters and frequency counts over sliding windows; still, their techniques rely on keeping simple equi-width counters within the sketch, and, thus, cannot provide any meaningful error guarantees, especially for small query ranges. Similarly, the hybrid histograms of Qiao et al. [24] combine exponential histograms with simplistic equi-width histograms for answering sliding-window range queries; again, these structures cannot give meaningful bounds on the approximation error and cannot be composed in a distributed setting.

## 3. PRELIMINARIES

ECM-sketches combine the functionalities of Count-Min sketches [10] and exponential histograms [12]. We now describe the two structures, focusing on the aspect related to our work.

**Count-Min Sketches.** Count-Min sketches are a widely applied sketching technique for data streams. A Count-Min sketch is composed of a set of $d$ hash functions, $h_1(\cdot), h_2(\cdot), \ldots, h_d(\cdot)$, and a 2-dimensional array of counters of width $w$ and depth $d$. Hash function $h_j$ corresponds to row $j$ of the array, mapping stream items to the range of $[1 \ldots w]$. Let $CM[i,j]$ denote the counter at position $(i, j)$ in the array. To add an item $x$ of value $v_x$ in the Count-Min sketch, we increase the counters located at $CM[h_j(x), j]$ by $v_x$, for $j \in [1 \ldots d]$. A point query for an item $q$ is answered by hashing the item in each of the $d$ rows and getting the minimum value of the corresponding cells, i.e., $\min_{j=1}^{d} CM[h_j(q), j]$. Note that hash collisions may cause estimation inaccuracies – only overestimations. By setting $d = \lceil \ln(1/\delta) \rceil$ and $w = \lceil e/\epsilon \rceil$, where $e$ is the base of the natural logarithm, the structure enables point queries to be answered with an error of less than $\epsilon ||a||_1$, with a probability of at least $1 - \delta$, where $||a||_1$ denotes the number of items seen in the stream. Similar results hold for range and inner product queries.

**Exponential Histograms.** Exponential histograms [12] are a deterministic structure, proposed to address the basic counting problem, i.e., for counting the number of true bits in the last $N$ stream arrivals. They belong to the family of methods that break the sliding window range into smaller windows, called buckets or basic windows, to enable efficient maintenance of the statistics. Each bucket contains the aggregate statistics, i.e., number of arrivals and bucket bounds, for the corresponding sub-range. Buckets that no longer overlap with the sliding window are expired and discarded from the structure. To compute an aggregate over the whole (or a part of) sliding window, the statistics from all buckets overlapping with the query range are aggregated. For example, for basic counting, aggregation is a summation of the number of true bits in the buckets. A possible estimation error can be introduced due to the oldest bucket inside the query range, which usually has only a partial overlap with the query. Therefore, the maximum possible estimation error is bounded by the size of the last bucket.

| Notation | Description |
|---|---|
| $N$ | Length of the sliding window, in time units or # arrivals |
| $h_i(\cdot)$ | Hash function $i$ of the Count-Min sketch |
| $a_r, b_r$ | Substream of stream $a, b$, within the query range $r$ |
| $f_a(x, r)$ | Frequency of item $x$ in stream $a$, within the query range $r$ |
| $E_a(i, j, r)$ | Estimated value of the ECM-sketch counter for stream $a$ in position $(i, j)$ for query range $r$ |
| $a_r \odot b_r, \widehat{a_r \odot b_r}$ | Real and estimated inner product of $a_r$ and $b_r$ |
| $u(N, S)$ | Upper bound of number of arrivals on stream $S$ within the sliding window of length $N$ |

**Table 1: Frequently used notation.**

To reduce the space requirements, exponential histograms maintain buckets of exponentially increasing sizes. Bucket boundaries are chosen such that the ratio of the size of each bucket $b$ with the sum of the sizes of all buckets more recent than $b$ is upper bounded. In particular, the following invariant (*invariant 1*) is maintained for all buckets $j$: $C_j / (2(1 + \sum_{i=1}^{j-1} C_i)) \leq \epsilon$ where $\epsilon$ denotes the maximum acceptable relative error and $C_j$ denotes the size of bucket $j$ (number of true bits arrived in the bucket range), with bucket 1 being the most recent bucket. Queries are answered by summing the sizes of all buckets that fully overlap the query range, and half of the size of the oldest bucket, if it partially overlaps the query. The estimation error is solely contained in the oldest bucket, and is therefore bounded by this invariant, resulting to a maximum relative error of $\epsilon$.

## 4. ECM-SKETCHES

We now describe ECM-sketches (short for Exponential Count-Min sketches), a composable sketch for maintaining data stream statistics over sliding windows in distributed environments. ECM-sketches combine the functionality of Count-Min sketches and sliding windows, and support both time-based and count-based sliding windows under the cash register model. Therefore, they can be used for compactly summarizing high-dimensional streams over sliding windows, i.e., to maintain the observed frequencies of the stream items within the sliding window range.

The core of the structure is a modified Count-Min sketch. Count-Min sketches alone cannot handle the sliding window requirement. To address this limitation, ECM-sketches replace the Count-Min counters with sliding window structures. Each counter is maintained as a sliding window, covering the last $N$ time units, or the last $N$ arrivals, depending on whether we need time-based or count-based sliding windows.

As discussed in Section 2, there have been several algorithms proposed for sliding window maintenance. Due to the large expected number of sliding window counters in ECM-sketches, we require an algorithm with a small memory footprint. Randomized sliding window synopses are therefore not a good choice. Instead, we employ exponential histograms [12], a compact and efficient deterministic synopsis. Each of the Count-Min counters is implemented as an exponential histogram, configured to provide an $\epsilon$ approximation for any query within a sliding window of length $N$, i.e., the estimation $\hat{x}$ of the counter for any query range within the sliding window length is in the range of $(1 \pm \epsilon)x$ of the true value $x$ of the counter. We will be discussing our choice for exponential histograms again in more detail in the following section, where we will consider alternative deterministic and randomized algorithms.

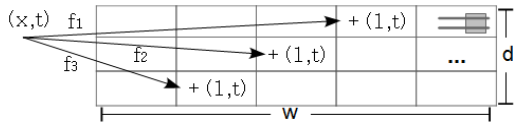

**Figure 1: Adding an element to the ECM-sketch.**

Adding an item $x$ to the structure is similar to the case of the standard Count-Min sketches. The process for time-based sliding windows is depicted in Figure 1. First, the counters $CM[h_j(x), j]$, where $j \in \{1 \ldots d\}$, corresponding to the $d$ hash functions are detected. For each of the counters, we register the arrival of the item at time $t$, and remove all expired information, i.e., the buckets of the exponential histogram that have no overlap with the sliding window range. The process for count-based sliding windows is similar, but instead of registering each arrival with system time $t$, we register it with the count of arrivals since the beginning of the stream.

The challenges that need to be addressed for the integration of exponential histograms with Count-Min sketches are: (a) to take into account the additional error introduced by the sliding window counters for deriving the accuracy guarantees for ECM-sketches (presented in the remainder of this section), and, (b) to enable composition of a set of ECM-sketches to a single ECM-sketch representing the order-preserving aggregation of the corresponding individual streams (Section 5).

## 4.1 Query Answering

We now explain how ECM-sketches support point queries, inner product queries, and self-join queries, and we derive probabilistic guarantees for the accuracy of the estimation. Our analysis covers both sliding window models, i.e., time-based and count-based.

**Point Queries.** A point query $(x, r)$ is a combination of an item identifier $x$, and the query range $r$ defined either as number of time units or number of arrivals. Point queries are executed as follows. The query item is hashed to the $d$ counters $CM[h_j(x), j]$ where ($j \in \{1 \ldots d\}$), and the estimate of each counter $E(h_j(x), j, r)$ for the query range is computed. The estimate value for the frequency of $x$ is $\hat{f}(x, r) = \min_{j=1 \ldots d} E(h_j(x), j, r)$.

Let $\delta_{cm}$ and $\epsilon_{cm}$ denote the configuration parameters of the Count-Min sketch, whereas $\epsilon_{sw}$ denotes the configuration parameter of the exponential histogram. With $||a_r||_1$ we denote the number of arrivals within the query range. The following theorem provides probabilistic guarantees for the approximation quality.

THEOREM 1. $|\hat{f}(x, r) - f(x, r)| \geq (\epsilon_{sw} + \epsilon_{cm} + \epsilon_{sw}\epsilon_{cm})||a_r||_1$ with probability at most $\delta = \delta_{cm}$.

PROOF. Special case of Theorem 3, proved in the appendix. □

As is typical for small-space sketches, the error guarantees are relative to the stream characteristics, i.e., the L1 norm. For all pairs of $\epsilon_{sw}$ and $\epsilon_{cm}$ satisfying $\epsilon_{sw} + \epsilon_{cm} + \epsilon_{sw}\epsilon_{cm} = \epsilon$, the maximum estimation error will be $\epsilon||a_r||_1$. (Note that $\epsilon \approx \epsilon_{cm} + \epsilon_{sw}$, since typically $\epsilon_{sw}, \epsilon_{cm} < 0.5$, and thus the product $\epsilon_{sw}\epsilon_{cm}$ is much smaller than the two linear terms.) The optimal pair of $\epsilon_{cm}$ and $\epsilon_{sw}$ is the one that minimizes memory utilization. The worst-case memory requirements of the structure are minimized as follows. The required memory per sliding window counter is $O(\frac{1}{\epsilon_{sw}} \log^2 Z)$, where $Z$ denotes the maximum possible count of each item in the sliding window. Therefore, the maximum required memory is mem $= \frac{c}{\epsilon_{sw}} \log^2 Z \times w \times d$, with $c$ denoting a constant, $w = \lceil e/\epsilon_{cm} \rceil$, and $d = \lceil \ln(1/\delta_{cm}) \rceil$. By derivation we find that the memory bound is minimized for $\epsilon_{sw} = \epsilon_{cm} = \sqrt{\epsilon + 1} - 1$, and becomes $O(\frac{\ln^2 Z \ln(1/\delta_{cm})}{\epsilon_{sw}\epsilon_{cm}}) = O(\frac{\ln^2 Z \ln(1/\delta_{cm})}{\epsilon})$.

**Inner Product and Self-Join Queries.** Another frequent query type is the cardinality of the inner product. Given two streams $a$ and $b$, the inner product is defined as $a \odot b = \sum_{x \in \mathcal{D}} f_a(x) \times f_b(x)$, where $\mathcal{D}$ denotes the input domain, i.e., the distinct input elements, and $f_a(x)$ (resp. $f_b(x)$) denotes the frequency of element $x$ in stream $a$ (resp. stream $b$). Self-join queries, also called the second frequency moment $F_2$, are a special case of inner product queries defined over a single stream: $F_2(a) = \sum_{x \in \mathcal{D}} (f_a(x))^2$. Both inner product queries and self-join queries are very important for databases, e.g., for building query execution plans, and they can be efficiently and accurately computed for streams with the cash register and turnstile model. However, similar to point queries, computing these queries over sliding windows is challenging.

ECM-sketches can be used to address this type of queries as well. Let $a_r$ (resp. $b_r$) denote the substream of stream $a$ (resp. $b$) within the query range. With $CM_a$ we denote the corresponding ECM-sketch for stream $a_r$, and with $E_a(i, j, r)$ we denote the estimated value of the counter of $CM_a$ in position $(i, j)$, for query range $r$. Also, $f_a(x, r)$ and $\hat{f}_a(x, r)$ denote the real and estimated frequency of $x$ in stream $a_r$.

The inner product of two streams $a$ and $b$ in a range $r$ is defined as $a_r \odot b_r = \sum_{x \in \mathcal{D}} f_a(x, r) f_b(x, r)$. Using the ECM-sketches of $a$ and $b$, we estimate it as follows: $\widehat{a_r \odot b_r} = \min_j (\widehat{a_r \odot b_r})_j$, where $(\widehat{a_r \odot b_r})_j = \sum_{i=1}^{w} E_a(i, j, r) \times E_b(i, j, r)$. The following theorem bounds the approximation error of this estimation.

THEOREM 2. $|\widehat{a_r \odot b_r} - a_r \odot b_r| \geq (\epsilon_{sw}^2 + 2\epsilon_{sw} + \epsilon_{cm}(1 + \epsilon_{sw})^2)||a_r||_1 ||b_r||_1$ with probability at most $\delta = \delta_{cm}$.

PROOF. In the appendix. □

The error is therefore $\approx (2\epsilon_{sw} + \epsilon_{cm})||a_r||_1 ||b_r||_1$, since the higher-order components are dominated by $\epsilon_{sw}$ and $\epsilon_{cm}$. Similar to the analysis for point queries, we can find the optimal pair of $\epsilon_{sw}$ and $\epsilon_{cm}$ guaranteeing a maximum error of $\epsilon||a_r||_1 ||b_r||_1$ by using derivation on the total memory requirements: $\epsilon_{sw} = -1 - \frac{3+3\epsilon}{3^{\frac{4}{3}} \left(9+9\epsilon+\sqrt{3}\sqrt{28+57\epsilon+30\epsilon^2+\epsilon^3}\right)^{\frac{1}{3}}} + \frac{\left(9+9\epsilon+\sqrt{3}\sqrt{28+57\epsilon+30\epsilon^2+\epsilon^3}\right)^{\frac{1}{3}}}{3^{\frac{2}{3}}}$ and $\epsilon_{cm} = \frac{\epsilon - \epsilon_{sw}^2 - 2\epsilon_{sw}}{(1+\epsilon_{sw})^2}$.

## 4.2 Extensions

### 4.2.1 Time-based Vs Count-based ECM-Sketches

Exponential histograms were originally developed for count-based sliding windows. They can be easily extended for time-based sliding windows as follows. First, each entry in the data structure is identified using its arrival time, instead of using its position in the stream. To reduce memory, arrival times are stored in wraparound counters of $O(\log(N))$ bits, where $N$ is the length of the sliding window, e.g., in milliseconds. Second, entries expire based on their arrival time, and not on their position in the stream. Finally, we require an upper bound of the number of arrivals within the sliding window time range for each stream $S$, denoted as $u(N, S)$. Note that this is required only for computing the maximum memory requirements of the structure a priori; it does not have an impact on the actual required memory or quality of ECM-sketches. Furthermore, the bound can be very loose without a noticeable change on the estimated space requirements, because space complexity increases only logarithmically with $u(N, S)$.

**Complexity.** We use $N$ to denote the length of the sliding window, either in number of arrivals or in time, depending on the desired sliding window model. With $u(N, S)$ we denote the upper bound of the number of arrivals in stream $S$ within a sliding window of length $N$. Also, $g(N, S) = \max(u(N, S), N)$.

To get an $\epsilon_{sw}$-approximation of the number of one-bits in the sliding window, exponential histograms require $O(\log(N)+$

| | Exponential Histogram | Deterministic Wave | Randomized Wave |
|---|---|---|---|
| Memory | $O\left(\frac{1}{\epsilon}\ln(\frac{1}{\delta})\ln^2(g(N,S))\right)$ | $O\left(\frac{1}{\epsilon}\ln(\frac{1}{\delta})\ln^2(g(N,S))\right)$ | $O\left(\frac{1}{\epsilon^2}\ln^2(\delta)\ln^2(u(N,S))\right)$ |
| Amort. update | $O(\ln(1/\delta))$ | $O(\ln(1/\delta))$ | $O(\ln^2(\delta))$ |
| Worst update | $O(\ln(1/\delta)\ln(u(N,S)))$ | $O(\ln(1/\delta))$ | $O(\ln^2(\delta)\ln(u(N,S)))$ |
| Query | $O(\ln(1/\delta)\ln(u(N,S))/\sqrt{\epsilon})$ | $O(\ln(1/\delta)\ln(u(N,S))/\sqrt{\epsilon})$ | $O(\ln^2(\delta)(\ln(u(N,S))+1/\epsilon^2))$ |

Table 2: Computational and space complexity of ECM-sketches. Function $g(N,S)$ is used as a shortcut for $\max(u(N,S),N)$.

$\log\log(u(N,S))$) memory per bucket, to store the bucket size and bucket boundaries. The number of buckets is $O(\log(u(N,S))/\epsilon_{sw})$, yielding a total memory of $O(\log^2(g(N,S))/\epsilon_{sw})$. With respect to computational cost, the update cost per element is $O(\log(u(N,S)))$ worst-case, and $O(1)$ amortized time. Queries covering the whole sliding window are executed in constant time. For queries with range $N' < N$, the required time is $O(\log(u(N,S)/\epsilon_{sw}))$. The extra time is required for finding the oldest bucket overlapping with the query, assuming sequential access. If the storage model of the buckets supports random access, e.g., a fixed-length array, then this time can be further reduced to $O(\log(\log(u(N,S)/\epsilon_{sw})))$, by employing binary search.

The space complexity of ECM-sketches is as follows. For the Count-Min array, we require an array of width $w = \lceil e/\epsilon_{cm}\rceil$ and depth $d = \lceil\ln(1/\delta)\rceil$. Each cell in the array stores an exponential histogram, requiring $O(\log^2(g(N,S))/\epsilon_{sw})$ bits. Therefore, the total memory requirements are $O(\frac{1}{\epsilon_{sw}\epsilon_{cm}}\log^2(g(N,S))\log(1/\delta))$. With respect to the time complexity, adding an element requires computing $d$ hash functions, and updating $d$ separate exponential histograms. The amortized complexity for each arrival is therefore $O(d) = O(\log(1/\delta))$, whereas the worst-case complexity is $O(d\log(u(N,S))) = O(\log(u(N,S))\log(1/\delta))$. Finally, query execution takes $O(\log(1/\delta))$ time for a query of range $N'$ equal to $N$. For $N' < N$, the execution cost is $O(d\log(u(N,S))/\epsilon_{sw}) = O(\log(1/\delta)\log(u(N,S))/\sqrt{\epsilon})$ with sequential access to buckets, e.g., using a linked list. With random access support, binary search can be used for finding the last relevant bucket for each query, reducing the query cost to $O(\log(1/\delta)\log(\log(u(N,S))/\sqrt{\epsilon}))$.

### 4.2.2 ECM-Sketches based on Waves

The sliding window counters can also be materialized using other sliding window algorithms. In the literature, two such algorithms are particularly well-known: (a) deterministic waves, and, (b) randomized waves [15]. We now show how ECM-sketches can incorporate these algorithms, and discuss the positive and negative aspects of each variant.

**Deterministic Waves.** Deterministic waves [15] have identical memory requirements with exponential histograms, and they outperform exponential histograms with respect to worst-case complexity for updates, requiring always constant time. As such, the space and computational complexity of ECM-sketches based on deterministic waves is the same to the one of sketches based on exponential histograms, with the only difference being the worst-case update complexity, which is $O(\log(1/\delta))$.

A downside of deterministic waves is that they require knowledge of the upper bound of the number of arrivals $u(N,S)$ during the initialization of the data structures, to decide on the required number of queues/levels. Any overestimation of $u(N,S)$ is therefore translated to an increase on the space requirements – logarithmic with $u(N,S)$. It is important to note that this constraint is substantially less limiting compared to the constraints of previous algorithms, e.g., [27], which required an upper bound for the *total number of items in all streams*, and therefore could not be applied to dynamic networks, with an unknown number of participating nodes and streams.

**Randomized Waves.** Randomized waves [15] provide an $(\epsilon,\delta)$ approximation for the basic counting problem, i.e., $Pr[|\hat{x}-x| \leq \epsilon_{sw}x] \geq 1-\delta_{sw}$, where $\hat{x}$ and $x$ denote the estimated and real number of true bits in the sliding window range respectively. This structure has substantially higher space complexity compared to the deterministic counterparts – $O(1/\epsilon_{sw}^2)$ instead of $O(1/\epsilon_{sw})$. However, randomized waves are important for distributed applications, as they enable lossless aggregation of individual summaries to a single summary corresponding to the aggregated data. Therefore, we also consider randomized waves for integration with the ECM-sketch.

The space complexity of ECM-sketches based on randomized waves is derived by multiplying the space complexity of the two basic structures: $O\left(\log(\delta_{cm})\log(\delta_{sw})\log^2(f(N,S))/(\epsilon_{cm}\epsilon_{sw}^2)\right)$. Inserting a new element requires $O(\log(\delta_{cm})\log(\delta_{sw}))$ amortized time, and $O(\log(\delta_{cm})\log(\delta_{sw})\log(f(N,S)))$ worst-case time. Finally, query execution takes $O(\log(\delta_{cm})\log(\delta_{sw})(\log(f(N,S))+1/\epsilon_{sw}^2))$ with sequential access to buckets and $O(\log(\delta_{cm})\log(\delta_{sw})(\log\log(f(N,S))+\log(1/\epsilon_{sw}^2)))$ time with random access.

THEOREM 3. $|\hat{f}(x,r)-f(x,r)| \geq (\epsilon_{sw}+\epsilon_{cm}+\epsilon_{sw}\epsilon_{cm})||a_r||_1$ with probability at most $\delta = \delta_{sw}+\delta_{cm}$.

PROOF. In the appendix. □

By derivation on the total memory usage, we can find the combination of $\epsilon_{sw}$ and $\epsilon_{cm}$ that minimizes the memory bound: $\epsilon_{sw} = \frac{\sqrt{\epsilon^2+10\epsilon+9}+\epsilon-3}{4}$ and $\epsilon_{cm} = \frac{3\epsilon-\sqrt{\epsilon^2+10\epsilon+9}+3}{\epsilon+\sqrt{\epsilon^2+10\epsilon+9}+1}$. The optimal space complexity becomes $O\left(\log(\delta_{cm})\log(\delta_{sw})\log^2(f(N,S))/\epsilon^2\right)$, and for $\delta_{cm}=\delta_{sw}=\delta/2$ it becomes $O\left(\log^2(\delta)\log^2(f(N,S))/\epsilon^2\right)$.

Table 2 summarizes the main results for the combination of ECM-sketches and the three sliding window structures. The results correspond to both time-based and count-based sliding windows.

## 5. ORDER-PRESERVING AGGREGATION

For many distributed applications, such as the network monitoring application described in the introduction, we require aggregating individual ECM-sketches $CM_1, CM_2, \ldots, CM_n$, each one corresponding to stream $S_1, S_2, \ldots, S_n$, to get a single ECM-sketch $CM_\oplus$ that corresponds to the logical stream $S_\oplus = S_1 \oplus S_2 \oplus \ldots \oplus S_n$. The $\oplus$ operator is defined as an aggregation that preserves the ordering and arrival time of the events. Standard Count-Min sketches allow this aggregation, as long as all sketches are constructed with identical dimensions and hash functions. For this, they rely on the linearity of the Count-Min counters, which are simple integers in the general case. However, this does not trivially hold for ECM-sketches, where the counters are not simple numbers but complex sliding window structures, since the analysis of exponential histograms (as well as all other deterministic sliding window structures), does not cover linearity. Although randomized structures cover linearity by default, these are substantially more expensive, and not preferable for ECM-sketches. Therefore, we now consider the order-preserving aggregation of deterministic sliding window structures. Note that this problem is interesting by itself, since these data structures are widely used in the literature for maintaining statistics over sliding windows. We then extend our results to cover aggregation of the ECM-sketches.

## 5.1 Aggregation of Exponential Histograms

Consider a set of exponential histograms $EH_1, EH_2, \ldots, EH_n$, summarizing time-based sliding windows. All are configured to cover a sliding window of $N$ time units. The aggregation operation is denoted with $\oplus$, i.e., $EH_\oplus = EH_1 \oplus EH_2 \oplus \ldots \oplus EH_n$. With $EH_i^j$ we denote bucket $j$ of $EH_i$, and $|EH_i^j|$ denotes the bucket size (number of true bits). By convention, buckets are numbered such that bucket 1 is the most recent. The ending time of the bucket is denoted as $e(EH_i^j)$. To ease exposition, we use $s(EH_i^j)$ to denote the starting time of the bucket, even though this is not explicitly stored in the buckets. By construction, the starting time of a bucket is equal to the ending time of the previous bucket, i.e., $s(EH_i^j) = e(EH_i^{j-1})$.

To construct $EH_\oplus$ our methodology considers the individual exponential histograms as logs. The general idea is to reconstruct $EH_\oplus$ by assuming that half of the elements arrive at the starting time of each bucket, and the other half at the ending time of the bucket. Precisely, let $\mathcal{B}$ denote the list containing all buckets of all sliding windows. We initialize an empty time-based exponential histogram with error $\epsilon'$, configured to keep the last $N$ time units, and a maximum of $\sum_{i=1}^{n} |EH_i|$ elements. For each bucket $\mathcal{B}[i] \in \mathcal{B}$, we simulate the insertion in $EH_\oplus$ of $|\mathcal{B}[i]|$ true bits. Half of the bits are inserted with timestamp $s(\mathcal{B}[i])$, and the other half at time $e(\mathcal{B}[i])$. The insertions are simulated in the order defined by the starting and ending timestamps of the buckets.

THEOREM 4. *Consider $n$ time-based exponential histograms $EH_1, EH_2, \ldots, EH_n$, initialized with error parameter $\epsilon$, and covering the same time range. The exponential histogram $EH_\oplus$ initialized with error parameter $\epsilon'$, and constructed with the proposed aggregation algorithm answers any query within its time range for the stream $S_\oplus$ with a maximum relative error of $(\epsilon + \epsilon' + \epsilon\epsilon')$.*

We will now give the intuition of the proof. The formal proof is presented in the appendix. Each exponential histogram $EH$ of stream $S$ configured with error parameter $\epsilon$ can be used to reconstruct an approximate stream $S'$, as follows: For each bucket $b$ in $EH$, add $|b|/2$ true bits in time $s(b)$, and $|b|/2$ true bits in time $e(b)$. We argue that answering any query with starting time $s_q$ within the range of $EH$ using the reconstructed stream $S'$ will result to a maximum relative error $\epsilon$. Let $b_j$ be the bucket s.t. $s(b_j) < s_q \leq e(b_j)$. Therefore, the accurate answer $x$ of the query for stream $S$ is bounded by $x \geq \sum_{i=1}^{j-1} |b_i| + 1$ and $x \leq \sum_{i=1}^{j-1} |b_i| + |b_j|$. By construction, the reconstructed stream will contain a total of $\sum_{i=1}^{j-1} |b_i| + |b_j|/2$ items with timestamp greater than or equal to $s_q$. Therefore, answering the query by counting the number of true bits in the reconstructed stream with timestamp after $s_q$ will have a maximum error of $\max(h - \sum_{i=0}^{j-1} |b_i| + |b_j|/2, \sum_{i=0}^{j-1} |b_i| + |b_j|/2 - l) = |b_j|/2$. By invariant 1 of exponential histograms, $|b_j|/2 \leq \epsilon(1 + \sum_{i=1}^{j-1} |b_i|) \leq \epsilon x$. Therefore, the maximum difference between the answer estimated by stream $S'$ and the correct answer $x$ will be less than or equal to $\epsilon x$.

Our aggregation algorithm is equivalent to reconstructing each stream $S_i'$ from exponential histogram $EH_i$, and using these to recreate an exponential histogram $EH_\oplus$. The reconstruction of stream $S'$ introduces a maximum relative error $\epsilon$, as explained above. Summarizing $S'$ with a new exponential histogram we get an additional error $\epsilon'$. However, $\epsilon'$ is relative on the answer provided by stream $S'$, and not by $S$. Therefore, the absolute error due to the exponential histogram summarization will be $\epsilon'x'$, where $x' \in (1 \pm \epsilon)x$ and $x$ denoting the accurate answer on $S_i$. Summing both errors, we get a total relative error of $\epsilon + \epsilon' + \epsilon\epsilon'$.

For the special case when $\epsilon' = \epsilon$, the maximum relative error becomes $2\epsilon + \epsilon^2$. Concerning space and computational complexity, $EH_\oplus$ behaves as a standard exponential histogram, and therefore has the same complexity as presented in [12]. □

**Multi-level Aggregation.** It is frequently desired to aggregate sliding windows in more than one levels. For example, consider a hierarchical P2P network, where each peer maintains its own exponential histogram, and pushes it to its parent for aggregation at regular intervals. Since the aggregated exponential histograms have the same properties as the individual exponential histograms (albeit with a higher $\epsilon$), the above analysis also supports iterative aggregation of exponential histograms.

There are two types of approximation error that influence the estimation of an aggregated exponential histogram. A possible approximation error, denoted as $\text{err}_1$, is introduced due to halving of the size of the last bucket of the aggregated exponential histogram. This error occurs only at query time, and is independent of the number of performed aggregations. Therefore, at a multi-level aggregation scenario this error does not need to be propagated at the intermediary exponential histograms. A second type of error, termed as $\text{err}_2$, occurs due to the inclusion (exclusion) of data that arrived before (after) the query starting time in buckets that are accounted (not accounted) in the query result.

It turns out that the error $\text{err}_2$ is additive at the worst case (in absolute value). For instance, in the lowest level (Level 0) of the hierarchy, aggregating two exponential histograms (all with relative error $\epsilon$), having a true number of bits (in a given query range) equal to $i_1$ and $i_2$, will result at a maximum value for $\text{err}_2 \leq \epsilon(i_1 + i_2)$. In Level 1, in addition to the previous possible errors, $\epsilon(i_1 + i_2) + \epsilon(i_3 + i_4)$ stream items may be incorrectly registered at the wrong side of the query start time. A recursive repetition for $h$ levels results to $\text{err}_2 \leq h\epsilon i$, where $i = \sum_j i_j$. The total absolute error (including $\text{err}_1$) then becomes $\text{err} = \text{err}_2 + \text{err}_1 \leq h\epsilon i + \epsilon(i + h\epsilon i)$, resulting to a maximum relative error of $h\epsilon(1 + \epsilon) + \epsilon$.

In many applications, the number of aggregation levels can be predicted, or even controlled when constructing the network topology. For example, consider DHT-based or hierarchical P2P topologies, which typically enable a balanced-tree access to the peers of height $h = \log(N)$, where $N$ is the number of nodes. In such systems, initializing the individual exponential histograms with error $\frac{\sqrt{1 + 2h + h^2 + 4h\epsilon} - 1 - h}{2h}$ yields an aggregated exponential histogram of relative error $\epsilon$. Naturally, this causes a slight inflation of the size of the sliding window, by $O(\log(N))$. However, even with this inflation, exponential histograms are – even for extremely large networks – substantially smaller and more efficient than randomized data structures that enable error-free aggregation in the expense of memory proportional to $O(1/\epsilon^2)$ (see also Section 5.2).

**Deterministic Waves.** The aggregation technique trivially extends for deterministic waves. Recall that each wave is composed of $l$ levels, each covering a different range. To perform the aggregation, we start from the lowest level $l - 1$, and switch to a higher level every $(1/\epsilon + 1)/2$ bits, i.e., when the first entry in the higher level has arrived before the next entry in the current level. Repeating the calculation of the error bounds for the aggregation of deterministic waves becomes straightforward when we notice that invariant 1 of the exponential histograms is also true for deterministic waves.

**Count-based Exponential Histograms.** Although exponential histograms cover both time-based and count-based sliding windows, aggregation of exponential histograms is specific for time-based sliding windows. Count-based sliding windows do not contain sufficient information for allowing order-preserving aggregation. Even storing the system-wide time of the buckets would not be sufficient to allow such an aggregation. To illustrate this limitation, consider the two count-based exponential histograms depicted in Fig. 2. For each bucket we store the bucket id, the size of the bucket, the bucket completion time and the total number of arrivals until that time. An arrival in count-based sliding windows might be a true or a false bit. An example query can then be: *how many true bits arrived in the last 100 system-wide arrivals*. If these 100 system-wide arrivals were read between time 19 and 20, then the correct answer would be 1. However, it is also possible that the last 100 system-wide arrivals have arrived between time 3 and time 20, in which case the correct answer could be anything between 2 and 9. The information contained in the two exponential histograms is not sufficient to estimate this type of queries, as it only allows us to preserve the order of the true bits, but looses the order of the false bits, which is also important. Therefore, given only the exponential histograms, it is not possible to aggregate them in a way that preserves the ordering of both true and false bits. Deterministic and randomized waves also have the same limitation when it comes to order-preserving aggregation of count-based sliding windows.

## 5.2 Aggregation of Randomized Waves

Randomized waves were proposed in [15] to address the problem of distributed union counting: *counting the number of 1's in the position-wise union of $t$ distributed data streams, over a sliding window*. However, the existing algorithm for utilizing more than one randomized waves does not consider aggregation of several waves, to generate a single wave. It assumes that the individual randomized waves can be stored and accessed any time, which is inconvenient for large networks. To eliminate this assumption we now propose a slight variation of their algorithm that can produce a single randomized wave out of a set of individual waves, with the same probabilistic accuracy guarantees as the individual waves.

Our algorithm simulates the construction of the aggregate randomized wave $RW_\oplus$ by using only the information included in the individual randomized waves. Consider a set $\mathcal{R}$ of randomized waves $RW_1, RW_2, \ldots, RW_n$, configured to store a sliding window of $N$ time units, with error parameters $\epsilon$ and $\delta$. The aggregate randomized wave $RW_\oplus$ is initialized with the same $\epsilon$ and $\delta$ parameters, for storing a maximum of $\sum_{i=1}^{n}|RW_i|$ events over $N$ time units. Each level $l$ of $RW_\oplus$ is then constructed by concatenating the corresponding level $l$ from all individual randomized waves, sorting all events based on the timestamp, and keeping the last $c/\epsilon^2$ events. Recall that the number of levels of individual randomized waves is determined based on the maximum number of events in the sliding window. Therefore, it may happen that $RW_\oplus$ has more levels than individual randomized waves. To populate the lower levels of $RW_\oplus$, we rehash the events populating the last level of each individual randomized wave, as proposed in [15] when merging different levels from randomized waves.

The process of query execution and the accuracy guarantees remain the same as for the standard randomized waves.

## 5.3 Composability of ECM-Sketches

Consider a set of ECM-sketches $CM_1, CM_2, \ldots, CM_n$ with identical dimensions and hash functions. The ECM-sketch $CM_\oplus$ with each counter set to the sum of all corresponding counters from the individual sketches (as defined by the $\oplus$ operator), summarizes the information found in the individual sketches:

$$CM_\oplus[j,k] = CM_1[j,k] \oplus CM_2[j,k] \oplus \ldots \oplus CM_n[j,k]$$

To bound the estimation error, we consider the two sources of error in the aggregated ECM-sketch. The error due to the Count-Min sketch $\epsilon_{cm}$ does not change, since it only depends on the dimensionality of the Count-Min array, which is fixed. However, the error due to sliding window estimations at each counter might change with each aggregation. Let $\epsilon'_{sw}$ denote the error produced by the aggregation of the corresponding Count-Min counters, as discussed in Sections 5.1 and 5.2. Recall that this error depends on the data structure used for maintaining the sliding window. Similar to the case of individual ECM-sketches, the total error is $\epsilon = \epsilon_{cm} + \epsilon'_{sw} + \epsilon_{cm}\epsilon'_{sw}$, with probability $1 - \delta_{sw} - \delta_{cm}$.

|                 | $EH_1$ |      | $EH_2$ |     |     |     |      |
|-----------------|--------|------|--------|-----|-----|-----|------|
| Bucket id       | 2      | 1    | 5      | 4   | 3   | 2   | 1    |
| Size            | 1      | 1    | 8      | 4   | 2   | 1   | 1    |
| Completion time | 3      | 20   | 3      | 5   | 10  | 15  | 19   |
| Arrivals        | 500    | 1000 | 900    | 950 | 980 | 990 | 1000 |

**Figure 2: An example why aggregating count-based exponential histograms is not possible.**

## 6. OTHER APPLICATIONS

In addition to point and inner product queries, ECM-sketches can also address more complex requirements. We now briefly discuss two such cases: (a) finding the frequent items, and, (b) continuous monitoring of the value of inner joins or point queries over distributed streams. Additional problems, such as computing quantiles or answering range queries over sliding windows, can also be addressed, e.g., by adapting the algorithms proposed for Count-Min sketches [10] to employ ECM-sketches instead.

### 6.1 Finding the Frequent Items

Consider a stream $S$ containing items from the universe $\mathbb{U}$. The straightforward solution for finding the frequent items in the sliding window is to execute $|\mathbb{U}|$ point queries on the ECM-sketch, one for each item in the universe, and retain only the items above the desired frequency threshold. However, this approach carries a computational complexity of $O(|\mathbb{U}| \times \ln(1/\delta))$ for executing all queries and detecting the frequent items, which is clearly prohibitive for streaming algorithms.

A more efficient algorithm based on range sums is proposed by Cormode et al. [10], and can be adapted to ECM-sketches for addressing the sliding-window requirements. The algorithm relies on group testing, for progressively reducing the domain of candidate frequent items, until only the truly frequent items remain. The basic idea is to create $\log(|\mathbb{U}|)$ ECM-sketches, denoted as $CM_0, CM_1, \ldots CM_{\log(|\mathbb{U}|)-1}$, to keep the number of occurrences of ranges of items. The $i$'th ECM sketch is used to maintain the range sum of the necessary dyadic ranges of length $2^i$ for covering $\mathbb{U}$. A new arrival $x \in \mathbb{U}$ is handled by adding $\lfloor x/2^i \rfloor$ to $CM_i$, for $0 \leq i < \log(|\mathbb{U}|)$. To detect the frequent items, we start with $CM_{\log(|\mathbb{U}|)-1}$, estimating the number of occurrences of the contained dyadic ranges. If any of the dyadic ranges has an estimated frequency less than the frequency threshold $\phi$, the whole dyadic range is ignored, as it cannot contain a frequent item. For all ranges with frequency surpassing $\phi$, the test continues recursively by breaking the range in two, and using the ECM-sketch of the lower level.

There are some interesting variants of the above problem, mostly relating to the way the threshold $\phi$ is expressed by the user. If $\phi$ is given as a minimum number of occurrences of each item, then no further computation is needed to determine which dyadic ranges are frequent and which are infrequent. However, it is often useful to express $\phi$ as the ratio of the number of occurrences of each item to the total number of arrivals within the sliding window. For time-based sliding windows, we can estimate the total number of arrivals by maintaining an additional sliding window, e.g., a deterministic wave, and using its lower bound. A better alternative that does not require additional memory is to use ECM-sketch $CM_0$ to estimate the total number of arrivals, by summing all counters in each row, and getting the average value. Although this approach has the same error bounds, in practice it offers better estimation accuracy than maintaining a single additional sliding window, since the errors coming from all counters in each row are usually canceled out.

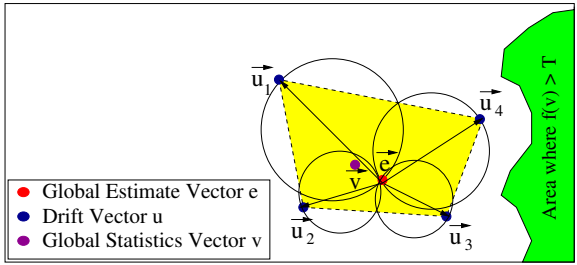

**Figure 3: Local constraints using the Geometric Approach. Each node constructs a sphere with diameter the drift vector u of the node and the estimate vector e. The global statistics vector v is guaranteed to lie in the convex hull of e, u1, u2, u3, u4. The union of the local spheres covers the convex hull.**

This estimation based on ECM-sketches may result to false positives and false negatives. Theorem 5 allows us to bound this error.

THEOREM 5. *The proposed algorithm uses $O((\log |\mathbb{U}|/\epsilon) \log(2 \log |\mathbb{U}|/(\delta\phi)) \log^2(g(N,S)))$ memory and amortized time $O(\log(2 \log |\mathbb{U}|/\delta) \log |\mathbb{U}|)$ per update, for detecting every item with frequency at least $(\phi+\epsilon)||a||_1$. With probability $1-\delta$, no item with frequency less than $\phi||a||_1$ is output.*

The same algorithm for approximating range sums can also be used for range queries, by noticing that all valid ranges within $\mathbb{U}$ can be expressed by a sum of dyadic ranges [10]. The error guarantees in this case are identical to the ones for Count-Min sketches, as described in [10], whereas the memory requirements are $O((1/\epsilon) \log(1/\delta) \log^2(g(N,S)) \log |\mathbb{U}|)$ bytes, for maintaining the $\log |\mathbb{U}|$ ECM-sketches.

## 6.2 Continuous Monitoring of Functions for Threshold Crossing

In many application domains, continuous monitoring of functions is required. ECM-sketches can also be used in these scenarios to reduce the memory and network requirements. We give the main intuition on how this can be done using self-join queries over sliding windows as an example.

We combine ECM-sketches with the geometric method [25]. The geometric method allows the distributed monitoring of complex (non-linear) functions defined over the average of local vectors (termed as *local statistics vectors*) maintained at sites. The goal is to to drastically reduce the required coordination for monitoring threshold crossing of such complex functions in a distributed network. The main idea is to distributively monitor the domain space where the average vector may lie. Each site monitors a portion of the corresponding subset of the domain space, with the corresponding monitoring zone often being expressed as a hypersphere. A common reference point of all such hyperspheres is the *global estimate vector*, which is the average vector computed during the last global communication (often called as a synchronization step) among all sites. Figure 3 depicts this process.

In this context, ECM-sketches are used to represent:

- The local statistics vectors at each site. The ECM-sketches are denoted as $\vec{sv}_1(t), \vec{sv}_2(t), \ldots, \vec{sv}_n(t)$, where $n$ is the number of sites. All sketches have an identical configuration.
- The global statistics vector. This vector is the current average over all local statistics vectors. The value of this vector is unknown to all sites, unless a synchronization takes place. The global statistics sketch is denoted as $\vec{sv}(t)$, and is computed by a linear aggregation of the local statistics sketches. We also use $\vec{se}(t)$ to denote the global estimate vector, which is the last known value of the global statistics vector.

Out of these two ECM-sketches, we can also compute the following two vectors, required by the geometric method:

- The statistics delta vectors, denoted using $\Delta\vec{sv}(t)$. This vector is equal to the difference between the local statistics vector and the corresponding vector that was transmitted in the last synchronization.
- The drift vectors, denoted as $\vec{su}_i(t)$, where $\vec{su}_i(t) = \vec{se}(t) + \Delta\vec{sv}_i(t)$. The global statistics vector is guaranteed to lie in the convex hull of the drift vectors, while this convex hull is covered by the union of hyperspheres monitored by the sites. Each hypersphere of a site is constructed with diameter the global estimate vector and the corresponding drift vector of the site [25].

To initialize the monitoring process, all nodes send their local statistics vectors $\vec{sv}_1(t), \vec{sv}_2(t), \ldots, \vec{sv}_n(t)$ to a coordinator. The coordinator aggregates all vectors using the algorithm for order-preserving aggregation of ECM-sketches, and computes a single global statistics vector $\vec{sv}(t)$. This global statistics vector is called the global estimate vector, and it is propagated to all network nodes, e.g., by using a hierarchy, or a broadcasting technique. This estimate vector is used by each participating node to extract a set of Count-Min sketches, one for each query range. Without loss of generality, assume that we have only a single query range, and $\vec{se}(t)$ denotes the corresponding extracted Count-Min sketch.

After each new arrival at time $t'$, node $p_i$ updates its local statistics vector $\vec{sv}_i$, and checks for a local constraint violation. For this check, $p_i$ extracts the statistics delta vector $\Delta\vec{sv}(t')$ from $\vec{sv}_i(t')$ as a Count-Min sketch, by querying each counter of $\vec{sv}_i(t')$ for its value within the time range $(t, t']$. By summing $\Delta\vec{sv}(t')$ with $\vec{se}(t)$ the node can compute the drift vector $\vec{su}_i(t')$, again as a Count-Min sketch, and construct the sphere of the geometric method. The sphere is formed with a center $\kappa = (\vec{se}(t) + \vec{su}_i(t'))/2$, and radius $\alpha = ||(\vec{se}(t) - \vec{su}_i(t'))||/2$. The geometric method guarantees that if the maximum and minimum value of the function within the sphere are at the same side of the threshold, then there can be no threshold crossing caused by this update. For computing the maximum and minimum value of the function efficiently, we currently have closed form equations for simple functions, like self-joins. Sharfman et al. [25] propose using numerical analysis algorithms, to compute these extrema, e.g., with Matlab. We are still working on this problem, to achieve efficient analytic solutions for more function types.

## 7. EXPERIMENTAL EVALUATION

Our experiments focused on evaluating ECM-sketches with respect to their scalability, effectiveness, and efficiency, as well as their suitability for distributed setups. The experiments were conducted using two frequently used real-life data sets, the world-cup'98 [2] (*wc'98*) and the Crawdad SNMP Fall 03/04 data set [21] (*snmp*). The wc'98 data set consists of all HTTP requests that were directed within a period of 92 days to the web-servers hosting the official world-cup 1998 website. It contains a total of 1.089 billion valid requests, served by 33 server mirrors. Each request was indexed using the web-page url as a key, i.e., the ECM-sketch could be used for estimating the popularity of each web-page. The snmp data set contains a total of 134 million records collected from the wireless network of Dartmouth college during the fall of 2003/2003. For this data set, we have used the (anonymised)

MAC addresses of the clients as keys for indexing. Therefore, the ECM-sketch enabled estimating the traffic volume generated by each user.

We have compared three sketch variants, differentiating on the employed sliding window algorithm: (a) the default variant described earlier which is based on exponential histograms, denoted as *ECM-EH*, (b) a variant using deterministic waves (*ECM-DW*), and, (c) a variant based on randomized waves (*ECM-RW*). The comparison between the variants was performed to demonstrate the influence of the sliding window algorithm to the performance of ECM-sketches.

## 7.1 Implementation Details

ECM-sketches were implemented in Java 1.7 using 32-bit addressing, and executed on a single idle core of an Intel Xeon 1.6 GHz machine. Deterministic and randomized waves were implemented as described in [15], including all optimizations. The queues were implemented as fixed-size deques. The waves were initialized using one event per millisecond as an upper bound for the number of arrivals within the sliding window. In practice, it is rarely possible to predict the maximum number of events per sliding window, and therefore conservative estimates, like this one, are often the only option. Concerning exponential histograms, [12] does not provide sufficient details for the implementation of the list of buckets. We therefore considered different possibilities for maintaining the buckets, including fixed arrays, deques, doubly-linked lists, and tree lists, and their combinations. The most efficient implementation was a combination of fixed arrays with deques, which enabled random access to buckets and constant-time bucket merges. Specifically, the bucket list was divided to different levels $L_0, L_1, \ldots, L_l$. Each level $L_i$ was initialized as a fixed-length deque, for storing only the buckets of size $2^i$. Furthermore, to save memory, all levels were initially set to null, and initialized on request. The space and computational complexity of our implementation is as described in Section 6, for the random-access model.

Unless otherwise noted, all ECM-sketches were set to monitor a sliding window of 1 million seconds (11.5 days). Queries were generated with an exponentially increasing range, i.e., query $q_i$ covered the range $[t-10^i, t]$, with $t$ denoting the time of the last arrival. For each range, a self-join query, as well as a set of point queries were constructed and executed. For thorough evaluation, we constructed one point query for each distinct item in the query range (i.e., estimating the popularity of each web-page in the wc'98 dataset, or the number of snmp messages generated by each MAC address in the snmp dataset).

## 7.2 Centralized Setup

In the centralized scenario, a single node monitors the whole stream and maintains an ECM-sketch, which is subsequently used for answering the queries. We first consider the tradeoff between memory requirements and estimation error. For this, we vary $\epsilon$ within the range of $[0.05, 0.25]$, keeping $\delta = 0.1$. For each $\epsilon$ value, we use the analysis presented in Section 4 to configure the ECM-sketch such that the required memory for the targeted query type is minimized – hence the difference in the cost of point queries and self-join queries for the same $\epsilon$ values.

Figures 4(a)-(d) plot the average and maximum observed error in correlation to the required memory for the two data sets. The figures are annotated with indicative $\epsilon$ values. The displayed error at the Y axis is relative to the number of events arriving within the query range, i.e., for point queries, err $= |\hat{f}(x,r) - f(x,r)|/||a_r||_1$ and for self-joins, err $= |\widehat{a_r \odot a_r} - a_r \odot a_r|/(||a_r||_1)^2$. Recall that the ECM-RW structure does not allow probabilistic guarantees for self-join queries, and is therefore not considered for this type of queries. Table 3 presents sample update rates for the considered variants, for $\epsilon = 0.1$.

Our first observation is that, for all variants, both the average and maximum observed errors are lower than the user-selected value $\epsilon$. However, the memory requirements of ECM-RW are at least an order of magnitude higher than the requirements of ECM-sketches based on the two deterministic structures for offering the same accuracy guarantees. As an example, for the wc'98 experiment with a moderate value of $\epsilon = 0.1$, the cost of maintaining the ECM-RW sketch is already 400 Mbytes, whereas the ECM-sketches based on exponential histograms and deterministic waves require less than a megabyte for satisfying the same guarantees (the simulation of ECM-RW configured with $\epsilon = 0.05$ could not be completed due to insufficient main memory). This happens because the memory requirements of randomized waves grow quadratically with $1/\epsilon$, whereas the two deterministic sliding window algorithms scale linearly. Note that this negative result applies to all known randomized sliding window algorithms, e.g., [27, 11], since they all scale quadratically with $1/\epsilon$. As such, ECM-sketches based on deterministic structures are more applicable for scenarios with non-specialized hardware, or hardware with less memory, like sensor networks and network devices. Comparing the two deterministic methods, we see that ECM-EH sketches are faster and more compact, requiring approximately half the space compared to the ones based on deterministic waves. All results are consistent for both data sets.

Summarizing, these results demonstrate that ECM-EH sketches are more efficient and compact compared to the other two variants, and that ECM-RW sketches require at least an order of magnitude more memory to satisfy the accuracy guarantees compared to the two variants based on deterministic sliding window structures.

## 7.3 Distributed Setup

The second series of experiments focused on evaluating the applicability of ECM-sketches for distributed setups. For this, we conducted simulations of distributed networks using the real-world distributions obtained from the two data sets. In particular, wc'98 contains the server identification for each of the 33 official world-cup servers answering the HTTP requests, whereas the records in the snmp data set contain the identification for each of the 535 monitored APs. For our simulations, these servers were organized in an architecture resembling a balanced binary tree of height $\lceil \log_2(n) \rceil$, where $n$ is the number of servers. All servers resided at the leaf nodes of the tree. Some of these servers were also randomly chosen to occupy the internal tree nodes, responsible for aggregation of the ECM-sketches coming from the children nodes. At the end of the aggregation process, the root node of the hierarchy was holding a single ECM-sketch, representing the order-preserving aggregation of the $n$ streams generated in $\lceil \log_2(n) \rceil - 1$ steps. ECM-DW sketches are not considered in this set of experiments, since they do not offer any advantages compared to ECM-EH sketches.

Figures 5(a)-(b) plot the average observed error for point and self-join queries in correlation to the network requirements for the whole aggregation to be completed. The results correspond to $\epsilon \in [0.05, 0.25]$ and $\delta = 0.1$. Note that the simulation with ECM-RW sketches did not complete for all $\epsilon$ values, due to insufficient memory resources at the machine simulating the $n$ nodes. To illustrate the accuracy loss due to this aggregation, Table 4 presents a comparison between the observed error of the centralized and the distributed ECM-sketches.

As expected, the process of iterative aggregations causes an increase of the observed error for ECM-EH sketches. This error however is still substantially lower than the upper bound derived by

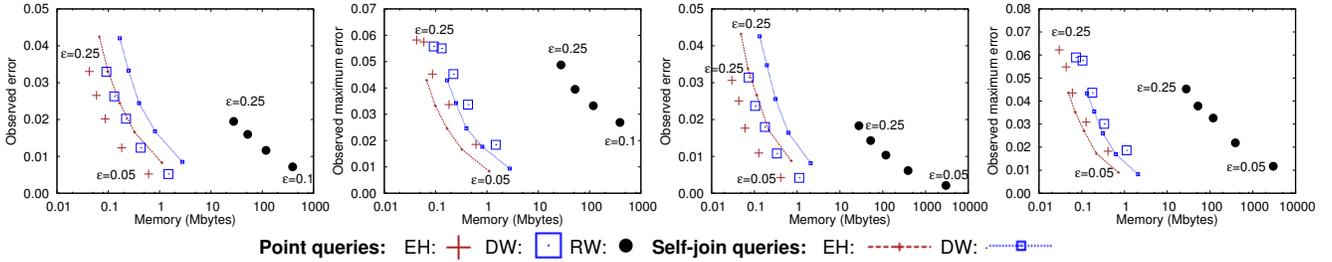

**Point queries:** EH: +    DW: □    RW: ●    **Self-join queries:** EH: ---+---    DW: ⋯□⋯

**Figure 4:** Average and maximum observed error in correlation to memory requirements for a centralized setup: (a)-(b) wc'98 data set, (c)-(d) snmp data set. The plots are annotated with indicative $\epsilon$ values.

|  | ECM-EH | ECM-DW | ECM-RW |
|---|---|---|---|
| wc'98 | 1486314 | 1167704 | 177149 |
| snmp | 736595 | 667036 | 105825 |

**Table 3:** Update rate (updates per second) for the centralized setup ($\epsilon = 0.1$).

| $\epsilon$ | Data set | Point queries ECM-EH Centr.:Distr. | Ratio | Self join ECM-EH Centr.:Distr. | Ratio | Point queries ECM-RW Centr.:Distr. | Ratio |
|---|---|---|---|---|---|---|---|
| 0.1 | wc'98 | 0.012:0.015 | 1.234 | 0.012:0.015 | 1.231 | 0.007:N/A | N/A |
| 0.2 | wc'98 | 0.027:0.031 | 1.164 | 0.026:0.029 | 1.131 | 0.016:0.016 | 1.008 |
| 0.1 | snmp | 0.011:0.011 | 1.042 | 0.010:0.011 | 1.021 | 0.006:0.006 | 1.031 |
| 0.2 | snmp | 0.025:0.026 | 1.037 | 0.025:0.025 | 1.016 | 0.014:0.014 | 0.986 |

**Table 4:** Observed error – loss is due to the iterative aggregation.

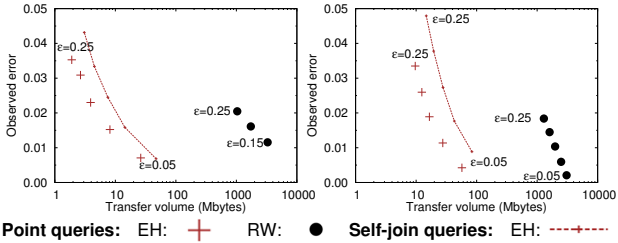

**Point queries:** EH: +    RW: ●    **Self-join queries:** EH: ---+---

**Figure 5:** Observed error in correlation to the network cost, for varying $\epsilon$: (a) wc'98 data set, (b) snmp data set.

the analysis. For example, for the case of the wc'98 data set with $\epsilon = 0.1$, the error bound is 0.3, whereas the average observed error after aggregation is less than 0.015, i.e., the increase due to the aggregation is less than 1/4 of the experimentally derived error of the centralized sketch. Concerning ECM-RW sketches, there is no systemic variation of the error, since randomized waves enable a lossless aggregation at the expense of a larger memory footprint. However, the network required for performing this aggregation using ECM-RW is higher by at least an order of magnitude compared to the transfer volume for the variant with exponential histograms. This requirement is prohibitive for a large set of application scenarios, like sensor and mobile networks, where high network usage causes battery drainage.

To further explore the influence of the network size on the estimation accuracy and network cost, we have also simulated an artificial network of $i$ servers, with $i = \{1, 2, 4, \ldots, 256\}$. The nodes were again placed as leaf nodes on a balanced binary tree, and the requests were divided uniformly across them. Figure 6(a) and (c) plot the average observed error in correlation to the network size, for $\epsilon = \delta = 0.1$. As expected, for ECM-EH sketches, increasing the number of nodes leads to a small increase on the observed estimation error. On the other hand, the aggregation process does not affect the accuracy of ECM-RW sketches, due to the lossless aggregation of randomized waves. However, the network cost for aggregating the sketches based on randomized waves (Figure 6(b) and (d)) is at least an order of magnitude higher compared to ECM-EH. This limits the applicability of ECM-sketches based on randomized waves to cases where a fast, fixed network is available, and makes the ability to merge deterministic sliding windows, e.g., based on exponential histograms, a very important contribution of this work.

Summarizing, this set of experiments showed that ECM-sketches based on exponential histograms can be aggregated with very small information loss. Compared to the lossless aggregation of ECM- sketches based on randomized waves, the sketches based on exponential histograms are substantially more compact, and are therefore applicable for a wider range of application scenarios, where network cost and memory is of the essence, such as P2P networks, sensor networks, and communication between network routers.

## 8. CONCLUSIONS

In this work we considered the problem of answering complex queries over distributed and high dimensional data streams, in the sliding window model. Our proposal, ECM-sketches, is a compact structure combining the state-of-the-art sketching technique for data stream summarization with deterministic sliding window synopses. The structure provides probabilistic accuracy guarantees for the quality of the estimation, for point queries and self-join queries, and can enable a broad range of problems, such as finding heavy hitters, computing quantiles, and answering range queries over sliding windows.

Focusing on distributed applications, we also showed how a set of ECM-sketches, each one representing an individual stream, can be aggregated to generate a single ECM-sketch that summarizes the stream produced by the order-sensitive aggregation of all individual streams. Interestingly, this is the first result in the literature enabling such aggregation for sketches that use deterministic sliding window synopses, and it is of high importance since deterministic synopses are generally a factor of $O(1/\epsilon)$ more compact than the best-known randomized synopsis for delivering an $\epsilon$-accurate approximation. In the same context, we demonstrated how ECM-sketches can be exploited for detecting frequent items, as well as within the geometric method for answering continuous queries.

ECM-sketches were thoroughly evaluated with a set of extensive experiments, using two large real-world datasets, and considering both centralized and distributed setups. The results verified the high performance of the structure. Compared to structures based on randomized sliding window synopses, ECM-sketches improve the memory and computational complexity by at least one order of magnitude. The same magnitude of improvement is observed with respect to the network requirements.

Our future work includes further investigation on employing ECM-sketches for the geometric method, for handling additional types of continuous queries over distributed sliding window streams.

**Acknowledgments.** This work was supported by the European Commission under ICT-FP7- LIFT-255951 (Local Inference in Massively Distributed Systems).

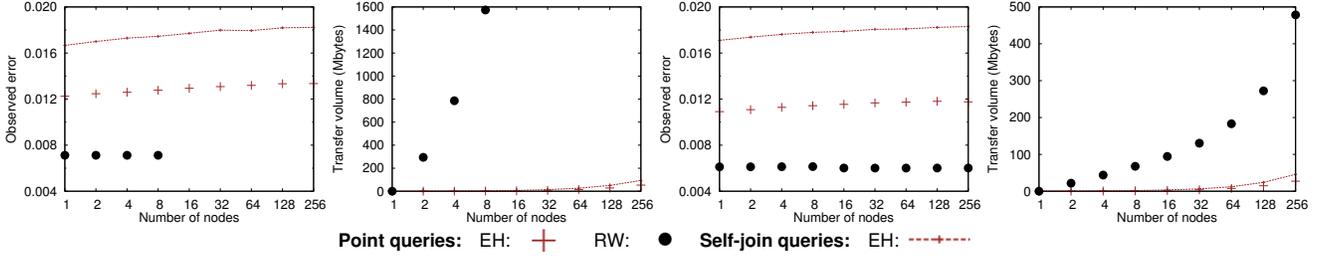

Figure 6: Observed error and network cost for different network sizes: (a)-(b) wc'98, (c)-(d) snmp.

## APPENDIX

PROOF OF THEOREM 2. We consider the estimation derived by any single row $j$ of the ECM-sketch. We first check the case of $E((\widehat{a_r \odot b_r})_j) > a_r \odot b_r$:

$$E((\widehat{a_r \odot b_r})_j - a_r \odot b_r) = \sum_{x \in \mathcal{D}} \hat{f}_a(x,r)\hat{f}_b(x,r) +$$

$$\sum_{\substack{p,q \in \mathcal{D}, p \neq q \\ h_j(p) = h_j(q)}} \hat{f}_a(p,r)\hat{f}_b(q,r) - \sum_{x \in \mathcal{D}} f_a(x,r)f_b(x,r)$$

$$\leq \sum_{x \in \mathcal{D}} f_a(x,r)f_b(x,r)(1+\epsilon_{sw})^2 +$$

$$\sum_{\substack{p,q \in \mathcal{D}, p \neq q \\ h_j(p) = h_j(q)}} f_a(p,r)f_b(q,r)(1+\epsilon_{sw})^2 - \sum_{x \in \mathcal{D}} f_a(x,r)f_b(x,r) =$$

$$(\epsilon_{sw}^2 + 2\epsilon_{sw})a_r \odot b_r + \sum_{\substack{p,q \in \mathcal{D}, p \neq q \\ h_j(p) = h_j(q)}} f_a(p,r)f_b(q,r)(1+\epsilon_{sw})^2 \quad (1)$$

From [10], we know that $E(\sum_{\substack{p,q \in \mathcal{D}, p \neq q \\ h_j(p) = h_j(q)}} f_a(p,r)f_b(q,r)) \leq \epsilon_{cm}||a_r||_1||b_r||_1/e$. Furthermore, by Markov inequality,

$$Pr[\forall j: \sum_{\substack{p,q \in \mathcal{D}, p \neq q \\ h_j(p) = h_j(q)}} f_a(p,r)f_b(q,r) \geq \epsilon_{cm}||a_r||_1||b_r||_1/e] \leq e^{-d} \leq \delta$$

Combining this with Eqn. 1, we get that with probability at least $1 - \delta$,

$$\widehat{a_r \odot b_r} - a_r \odot b_r \leq (\epsilon_{sw}^2 + 2\epsilon_{sw})a_r \odot b_r + \epsilon_{cm}(1+\epsilon_{sw})^2||a_r||_1||b_r||_1$$

Repeating the analysis for the case of $E((\widehat{a_r \odot b_r})_j) < a_r \odot b_r$ we get the following probabilistic guarantees:

$$a_r \odot b_r - \widehat{a_r \odot b_r} \leq (\epsilon_{sw}^2 + 2\epsilon_{sw})a_r \odot b_r$$

The bounds follow directly by noticing that $a_r \odot b_r \leq ||a_r||_1||b_r||_1$. □

PROOF OF THEOREM 3. By the estimation algorithm we know that there exists at least one row $j$, for which $E(h_j(x), j, r) = \hat{f}(x,r)$. Let us focus now on this row. We initially assume that we have an accurate algorithm to maintain the sliding window counters, i.e., errors are only due to hashing collisions. With $R(h_j(x), j, r)$ we denote the accurate number of bits that were added in the counter $(h_j(x), j)$, within the query range $r$. Note that, because of hashing collisions, the value of $R(h_j(x), j, r)$ might be greater than the real frequency of $x$, denoted as $f(x,r)$. In fact, since the counters are assumed to be accurate, the standard analysis introduced for count-min sketches may be applied. Therefore, $Pr[R(h_j(x), j, r) - f(x,r) \leq \epsilon_{cm}||a_r||_1] \geq 1 - \delta_{cm} \Rightarrow Pr[R(h_j(x), j, r) \leq f(x,r) + \epsilon_{cm}||a_r||_1] \geq 1 - \delta_{cm}$.

However, in practice, the sliding window algorithm may introduce errors to the computation of $R(h_j(x), j, r)$. Since all considered algorithms are $(\epsilon, \delta)$-approximate, we know that their estimation $E(h_j(x), j, r)$ has the following property: $Pr[|E(h_j(x), j, r) - R(h_j(x), j, r)| \leq \epsilon_{sw} R(h_j(x), j, r)] \geq 1 - \delta_{sw}$.

For the case that $E(h_j(x), j, r) > R(h_j(x), j, r)$, we have $Pr[E(h_j(x), j, r) \leq (1 + \epsilon_{sw}) R(h_j(x), j, r)] \geq 1 - \delta_{sw}$. Considering the two results together, we get:

$$Pr[E(h_j(x), j, r) \leq (1 + \epsilon_{sw}) R(h_j(x), j, r)] \geq 1 - \delta_{sw} \Rightarrow$$
$$Pr[E(h_j(x), j, r) \leq (1 + \epsilon_{sw})(f(x,r) + \epsilon_{cm}||a_r||_1)]$$
$$\geq 1 - \delta_{sw} - \delta_{cm} \Rightarrow$$
$$Pr[\hat{f}(x,r) - f(x,r) \leq \epsilon_{sw} f(x,r) + \epsilon_{cm}||a_r||_1 +$$
$$\epsilon_{cm}\epsilon_{sw}||a_r||_1] \geq 1 - \delta$$

Note that $\epsilon_{sw} f(x,r) + \epsilon_{cm}||a_r||_1 + \epsilon_{cm}\epsilon_{sw}||a_r||_1 \leq (\epsilon_{sw} + \epsilon_{cm} + \epsilon_{cm}\epsilon_{sw})||a_r||_1$. Therefore,

$$Pr[\hat{f}(x,r) - f(x,r) \leq (\epsilon_{sw} + \epsilon_{cm} + \epsilon_{cm}\epsilon_{sw})||a_r||_1] \geq 1 - \delta \quad (2)$$

With a similar analysis, the case of $E(h_j(x), j, r) < R(h_j(x), j, r)$ gives a much tighter constraint:

$$Pr[f(x,r) - \hat{f}(x,r) \leq \epsilon_{sw} f(x,r)] \geq 1 - \delta_{sw} \quad (3)$$

Note that the events considered by equations 2 and 3 are mutually exclusive. The proof is completed by taking the minimum of $Pr[f(x,r) - \hat{f}(x,r) \leq \epsilon_{sw} f(x,r)]$ and $Pr[\hat{f}(x,r) - f(x,r) \leq (\epsilon_{sw} + \epsilon_{cm} + \epsilon_{cm}\epsilon_{sw})||a_r||_1]$. □

PROOF OF THEOREM 4. We argue that $EH_\oplus$ approximates the exponential histogram of the logical stream, with a maximum relative error of $(1 + \epsilon)\epsilon' + \epsilon$, where $\epsilon$ is the error parameter of the initial exponential histograms. Consider a query for the last $q$ time units. With $s_q = t - q$ we denote the query starting time. Let $Q$ denote the index of the bucket of $EH_\oplus$ which contains $s_q$ in its range, i.e., $s(EH_\oplus^Q) \leq s_q \leq e(EH_\oplus^Q)$. With $i$ and $\hat{i}$ we denote the accurate and estimated number of true bits in the query range. According to the estimation algorithm, the estimation for the number of true bits in the stream will be $\hat{i} = 1/2|EH_\oplus^Q| + \sum_{1 \leq Y < Q}|EH_\oplus^Y|$. This estimation may be influenced by two types of approximation errors: (a) a possible approximation error of the overlap of bucket $EH_\oplus^Q$ with the query range, denoted as err$_1$, and, (b) a possible approximation error of $i$, denoted as err$_2$, because of the inclusion of data that arrived before $s_q$ in buckets $Y \leq Q$, or data that arrived after $s_q$ in buckets $Y > Q$. Let us now look into these two errors in more details.

With respect to err$_2$, recall that the contents of individual buckets are inserted to $EH_\oplus$ using the starting time and the ending time of the buckets. Therefore, it may happen that some bits arrive before $s_q$ but are inserted to $EH_\oplus$ with a timestamp after $s_q$, creating 'false positives'. The opposite is also possible. These bits are called out-of-order bits with respect to $s_q$. Clearly, out-of-order bits may lead to underestimation or overestimation of the query answer. The following lemma allows us to upper bound the number of out-of-order bits, and thereby control the maximum error err$_2$.

LEMMA 1. *Consider an (individual) exponential histogram $EH_z$ of stream $Z$, configured with error parameter $\epsilon$. The out-of-order bits with respect to the query starting time $s_q$ that $EH_z$ can generate are at most $\epsilon i_z$, with $i_z$ denoting the number of true bits arriving after $s_q$ in $Z$.*

PROOF. Due to the non-decreasing nature of bucket timestamps, there can be only one bucket with a start time less than $s_q$ and end time greater than or equal to $s_q$. Let this bucket be $EH_z^j$. All other buckets have both starting and ending time at the same side of $s_q$, and therefore their contents are always inserted with a timestamp at the correct side of $s_q$ and do not create out-of-order bits.

Since the ending time of $EH_z^j$ is at or after $s_q$, its most recent true bit has arrived at or after $s_q$, and should be included in the query range. Therefore, the number of true bits arriving at or after $s_q$ in stream $Z$ is $i_z \geq 1 + \sum_{b=1}^{j-1}|EH_z^b|$. Furthermore, since half of the bits of $EH_z^j$ are inserted using the ending time and half using the starting time of the bucket, the maximum number of out-of-order bits is $|EH_z^j|/2$. By construction (invariant 1):

$$\frac{|EH_z^j|}{2(1 + \sum_{b=1}^{j-1}|EH_z^b|)} \leq \epsilon \Rightarrow \frac{|EH_z^j|}{2} \leq \epsilon(1 + \sum_{b=1}^{j-1}|EH_z^b|) \leq \epsilon i_z \quad \square$$

The following lemma extends this result to all exponential histograms constituting $EH_\oplus$, for computing the total value of err$_2$:

LEMMA 2. *Consider the exponential histogram $EH_\oplus$, constructed by aggregating exponential histograms $EH_1$, $EH_2$, ..., $EH_n$. The maximum value of err$_2$ is $\epsilon i$, with $i = \sum_{x=1}^{n} i_x$ denoting the number of true bits that arrived in all streams during or after $s_q$.*

PROOF. Let err$_2(x)$ denote the number of out-of-order bits of stream $x$ with respect to $s_q$. Furthermore, $j_x = \max\{b|e(EH_x^b) \geq s_q\}$.

Notice that err$_2(x)$ is upper-bounded by Lemma 1. Due to the aggregation algorithm, err$_2 = \sum_{x=1}^{n}$err$_2(x)$. Observing that $\epsilon$ is the same across all EH, we have: err$_2 = \sum_{x=1}^{n}$err$_2(x) \leq \epsilon \sum_{x=1}^{n} i_x \leq \epsilon i$. □

Underestimation or overestimation of the overlap may also happen because of the halving of the size of bucket $EH_\oplus^Q$ during query time (err$_1$). As shown in [12], this process may introduce a maximum relative error of $\epsilon r$, where $r$ is the sum of the sizes of all buckets in $EH_\oplus$ with an index lower than $Q$ (i.e., with a starting time at least equal to $s_q$). Recall that $r$ may also include bits that arrived before $s_q$, which can however be upper bounded by Lemma 2. Therefore, the maximum underestimation or overestimation error is err$_1 = \epsilon' r \leq \epsilon'(i + \epsilon i) = \epsilon' i + \epsilon \epsilon' i$, with $i = \sum_{x=1}^{n} i_x$.

Summing err$_1$ and err$_2$, we get a maximum relative error of $(\epsilon + \epsilon' + \epsilon \epsilon')$. Theorem 4 follows directly. □